\documentclass[showpacs]{interact}

\usepackage{epstopdf}% To incorporate .eps illustrations using PDFLaTeX, etc.
\usepackage{subfigure}% Support for small, `sub' figures and tables

\usepackage[numbers,sort&compress,merge]{natbib}% Citation support using natbib.sty

\usepackage{soul}

\theoremstyle{plain}% Theorem-like structures

\theoremstyle{definition}

\theoremstyle{remark}

\usepackage{color,soul}
\begin{document}

\title{Adiabatic Quantum Estimation: A Numerical Study of the Heisenberg XX Model with Antisymmetric Exchange }

\author{
\name{L. Fathi Shadehi\textsuperscript{a}, H. Rangani Jahromi\textsuperscript{b}\thanks{Corresponding author: H. Rangani Jahromi. Email: h.ranganijahromi@jahromu.ac.ir}, and M. Ghanaatian\textsuperscript{a}}
\affil{\textsuperscript{a}Department of Physics, Payame Noor University (PNU), P.O. Box 19395-3697, Tehran, Iran; \textsuperscript
{b}Physics Department, Faculty of Sciences, Jahrom University, P.B. 7413188941, Jahrom, Iran.}
}

\maketitle

\begin{abstract}
In this paper, we address the adiabatic technique for quantum estimation of the azimuthal orientation of a  magnetic field.  Exactly solving
a model consisting of  a two-qubit system, where one of
which is driven by a static magnetic field  while the other is coupled with the magnetic field rotating adiabatically, we obtain the analytical expression of the quantum Fisher information (QFI). We investigate how the two-qubit system can be used to probe the azimuthal direction of the  field and analyze the roles of the intensities of the magnetic fields,  Dzyaloshinskii-Moriya interaction,  spin-spin coupling coefficient, and the polar orientation of the rotating field  on the  precision of the estimation. In particular, it is illustrated that the QFI trapping or saturation may occur if the qubit is subjected to a strong rotating field. Moreover, we discuss how the azimuthal direction of the rotating field can be estimated using only the qubit not affected by  that field and investigate the conditions under
which this strategy is more efficient than use of  the qubit locally interacting with the adiabatically rotating field. Interestingly, in the one-qubit scenario,   it was found  that when the rotating field
is weak, the best estimation is achieved by subjecting the probe to a static magnetic field.
\end{abstract}


\section{Introduction}
\par
Estimating unknown parameters of a quantum system is a necessary task for almost all branches of science and technology. Evidently,  any estimation would be associated with errors because of various factors such as natural stochasticity of the event in question or  imperfection of measurement devices, and hence estimated values are usually inaccurate.  
In the process of  estimating an unknown parameter of a quantum system, the quantum Fisher information (QFI) \cite{Braunstein34391994,Braunstein1351996,RanganiQIP4,RanganiIJMO} is a pivotal quantity depending on the state of the system and its derivate with respect to the unknown parameter. In fact, the QFI,  a fundamental notion in quantum metrology \cite{Giovannetti13302004,Giovannetti0104012006,RanganiAOP,RanganiAOP2,RanganiOPTC,RanganiJMO}, plays an important  role
in quantum detection because it provides a bound to the accuracy of quantum estimation.
The QFI, originally introduced by
Fisher \cite{Fisher7001925} and  defined by the Cram\'{e}r-Rao bound \cite{Cover2006}, is an extension of the Fisher information (FI) in the quantum regime. The FI concept, originating from the statistics, quantifies the accuracy of 
estimating the parameters and characterizes the optimal rate at which the neighbouring states may be distinguished
by measurement.  On the other hand, the QFI is defined by maximizing the FI over all possible positive operator valued measurements (POVM).

\par
A quantum probe is usually a  microscopic physical system prepared in a quantum superposition.  Therefore, the system may become very sensitive to decoherence effects and, in particular, to fluctuations  affecting  one  or  more  parameters which should be estimated \cite{RanganiQIP2019}. Hence, quantum sensing \cite{Degen0350022017} is  the art of exploiting the inherent fragility of quantum systems  to design the best  quantum  protocols  for  metrological  tasks. Up to now,   it has been proved that the quantum probes are useful in various branches of quantum metrology such as gravitational wave detectors \cite{McKenzie2311022002}, interferometry \cite{Lee23252002}
frequency spectroscopy \cite{BollingerR46491996}, atomic clocks \cite{Valencia26552004}, thermometry \cite{Guo0521122015,Correa2204052015, Kiilerich0421242018,RanganiQIP3,Ranganiscripta},  and even in sensing the magnetic
field \cite{ Taylor8102008,Degen2431112008,Matsuzaki0121032011, Jensen1608022014,Tanaka1708012015,Guo332542016,Ghirardi0121202012,Troiani2605032018, Danilin292018} (avian magnetoreception \cite{Maeda3872008,Pauls0627042013,Cai2305032013,Bandyopadhyay1105022012,Tiersch45172012,Mouloudakis0224132017,Guo58262017,RanganiHaseli} essentially belongs to this field).

  \par

The exact measurement of magnetic fields has many applications, from archaeology to the monitoring of brain activity,
  the local mapping of fields in a nanodevice and submarine detection \cite{Ripka1291992,Hamalainen4131993,Sander9812012,Le Sage2013,Jensen296382016}. One of the  most important applications  of magnetic field sensors  has  been found  in  the  area  of  biomagnetism \cite{Hamalainen4131993,Rodriguez4301999} defined as detecting  the  weak  magnetic  fields  produced  by  the  human organs such as brain and heart. In particular, measurements of the magnetic field originated from the  brain are applied to diagnose epilepsy, and to investigate neural responses to auditory and visual stimuli.
  On the other hand, 
  the highest sensitivities are obtained by superconducting
quantum interference devices \cite{Swithenby8011980} utilizing large atomic ensembles or macroscopic pick-up coils    
\cite{Wasilewski1336012010,Kominis5962003}, at the expense of detection volumes in
the $ cm^{3} $ range. A much higher spatial resolution, but less sensitivity,
is achieved by nitrogen-vacancy-centre sensors \cite{Taylor8102008} or atoms \cite{Wildermuth4352005,Vengalattore2008012007,Ockeloen1430012013,Mussel1030042014}, ions \cite{Ruster0310502017} or solid-state nanodevices \cite{Bal13242012}, ultimately at the single-particle level, as shown in pioneering important experiments \cite{Balasubramanian6482008,Maze6442008,Maiwald5512009,Baumgart2408012016}.

\par
In this paper, we  apply a two-qubit system to  probe the azimuthal direction of a magnetic field adiabatically rotating.
We exactly solve the Heisenberg   model, a simple spin chain model utilized to simulate many physical quantum systems such as quantum dots \cite{Trauzettel1922007}, nuclear spins \cite{Kane1331998}, superconductors \cite{Senthil42451999}, and optical lattices \cite{Sorensen22741999}. Moreover, Heisenberg model is ideal for generation of qubit states, thus attracting attention recently for realization of solid-based quantum computers. In \cite{Ozaydin522019}, the authors studied the  precision of phase estimation
	with thermal entanglement of two spins in Heisenberg XX model, in the presence of Dzyaloshinskii–
	Moriya  interaction. Moreover, considering a ferromagnetic XXZ spin-1/2 spin chain in the external field, the authors of  \cite{Rams0210222018} addressed the question of whether the super-Heisenberg scaling for quantum estimation is indeed
	realizable. In particular, they provided an experimental
	proposal of realization of the considered model via mapping the system to ultracold bosons in a periodically
	shaken optical lattice. Besides, it has been discussed how the 
	geometrical approach to quantum phase transition can improve
	estimation strategies for experimental inaccessible parameters in quantum Ising and Heisenberg X-Y models with an external field 
	 \cite{Invernizzi0421062008,Invernizzi1982010,Garnerone0572052009,Salvatori0221112014}.
 A steady state in the presence of a correlated dissipative Markovian noise to estimate
an unknown magnetic field has been studied in \cite{Guo332542016}. Here we show that the adiabatic state is another rich resource for the parameter estimation.

This paper is organized as follows: In Sec. II we give a brief description about the quantum adiabatic theorem and the QFI. The physical model is presented in Sec. III. We study the different scenarios for estimating the  magnetic field direction in Sec. IV. Finally, in  Sec. V,  the main results are summarized.

\section{PRELIMINARIES  \label{Preliminaries }}  
\subsection{The Adiabatic Approximation in Closed Quantum Systems \label{QAT}}
Let us start by
reviewing the adiabatic approximation in closed quantum
systems, subject to unitary evolution \cite{Sarandy0123312005}. In this situation, the evolution is governed by the time-dependent Schr\"{o}dinger equation

\begin{equation}\label{Shrodinger}
   H(t)|\psi(t)\rangle=i|\dot{\psi}(t)\rangle,
   \end{equation}
in which $  H(t) $ represents the Hamiltonian and $ |\psi(t)\rangle $ denotes a quantum
state in a d-dimensional Hilbert space. For simplicity, it is assumed that 
$  H(t) $ spectrum  is entirely discrete and
nondegenerate. Therefore, one can define an instantaneous basis of
eigenenergies by

\begin{equation}\label{Eigen1}
   H(t)|\xi_{j}(t)\rangle=\xi_{j}(t)|\xi_{j}(t)\rangle,
   \end{equation}
where the set of eigenvectors $|\xi_{j}(t)\rangle $ is orthonormal.
In this simplest case, because there is a one-to-one correspondence between energy levels  and eigenstates, the adiabaticity can be then defined as
the regime associated with an independent evolution of the
instantaneous eigenstates of $H(t)$, leading to level-anti-crossing phenomenon and continuous evolution of the instantaneous eigenstates at one time  to the
corresponding eigenstates at later times. In fact, when the system begins its evolution in a special eigenstate $ |\xi_{j}(0)\rangle $, then
it evolves to the instantaneous eigenstate $ |\xi_{j}(t)\rangle $ at a later
time $ t $, without any transition to the other energy levels \cite{RanganiPRSA}. A  general validity condition for adiabatic behaviour is given by
\begin{equation}\label{Adiabaticcondition}
   \max_{0\leq t\leq T} \bigg|\dfrac{\langle \xi_{k}|\dot{H}|\xi_{j}\rangle}{\xi_{j}-\xi_{k}}\bigg|\ll \min_{0\leq t\leq T} \big|\xi_{j}-\xi_{k}\big|,
   \end{equation}
 where $ T $ is the total evolution time. In the case of the degenerate spectrum, each degenerate eigenspace
of $ H(t) $, instead of individual eigenstates, has independent
evolution, whose validity conditions given by Eq. (\ref{Adiabaticcondition}) should 
be considered over eigenvectors with distinct energies. 

\subsection{Quantum Fisher Information\label{QFI}}
A standard scenario in quantum metrology may be described as follows: Firstly, the probe system, prepared in an appropriate initial state,
undergoes an evolution  imprinting the parameter information onto the evolved state, say
$ \rho(\lambda) $, and finally  a POVM measurement is performed on the probe.
Repeating the overall process  $ N $ times,  one can infer 
parameter $ \lambda $ from the statistics of the measurement outcomes by choosing an unbiased estimator. 
The accuracy of estimating $ \lambda $ is lower bounded by the quantum Cram\'{e}r-Rao inequality:

\begin{equation}\label{Cramer}
   \Delta \hat{\lambda}\geq \dfrac{1}{\sqrt{NF_{Q}(\lambda)}},
   \end{equation}
where the QFI associated with   unknown parameter $ \lambda $ encoded in quantum state $ \rho\left(\lambda \right) $ is defined as \cite{C. W. Helstrom,Braunstein1351996}
 \begin{equation}\label{01}
   F_{Q}\left( \lambda\right)=\text{Tr}\left[\rho\left(\lambda \right)L^{2} \right]=\text{Tr}\left[\left( \partial_{\lambda}\rho\left(\lambda \right)\right) L\right], 
   \end{equation}
in which $ L $, denoting the symmetric logarithmic derivative (SLD), is given by $ \partial_{\lambda}\rho\left(\lambda \right)=\frac{1}{2}\left(L\rho\left(\lambda \right)+\rho\left(\lambda \right)L\right), $ with $ \partial_{\lambda}=\partial/\partial\lambda $. Utilizing the spectrum decomposition of $ \rho\left(\lambda \right) $, $ \rho\left(\lambda \right)=\sum_{i}p_{i}|\phi_{i}\left\rangle \right\langle \phi_{i}|$, where $ |\phi_{i}\rangle $ and $ p_{i} $ represent the eigenvectors and eigenvalues of the matrix $ \rho\left(x \right) $, respectively; we can rewrite the QFI as follows \cite{Paris1252009}
\begin{eqnarray}\label{02}
\nonumber F_{Q}\left( \lambda\right)=\sum_{i,j}\frac{2}{p_{i}+p_{j}}|\langle \phi_{i}|\partial_{\lambda}\rho\left(\lambda \right)|\phi_{j}\rangle|^{2}~~~~~~~~~~~~~~~~~\\\
=\sum_{i} \frac{(\partial_{\lambda}p_{i})^{2}}{p_{i}}+2 \sum_{i\neq j} \frac{(p_{i}-p_{j})^{2}}{p_{i}+p_{j}} |\langle\phi_{i}|\partial_{\lambda} \phi_{j} \rangle|^{2},
\end{eqnarray} 
In the case of a pure quantum statistical model, i.e. $ \rho(\lambda)=|\psi_{\lambda}\rangle\langle \psi_{\lambda}| $, it is
possible to find  following simple expression \cite{Haine2304042016},

 \begin{equation}\label{PUREQFI}
   F_{Q}\left( \lambda\right)=4\big[\langle \partial_{\lambda}\psi_{\lambda}|\partial_{\lambda}\psi_{\lambda} \rangle-|\langle \psi_{\lambda}|\partial_{\lambda}\psi_{\lambda} \rangle|^{2}\big].
   \end{equation}
where $ \partial_{\lambda}\equiv \dfrac{\partial}{\partial \lambda} $.

\section{Physical model \label{Model}}

The Hamiltonian describing  two spin-1/2 Heisenberg XX model with 
antisymmetric super-exchange  Dzyaloshinskii-Moriya (DM) interaction and 
 in the presence of external magnetic fields takes the form

\begin{equation}\label{key1}
H(t)=J\left( {{\sigma }_{1}}^{x}\sigma _{2}^{x}+\sigma _{1}^{y}\sigma _{2}^{y} \right)+{{\vec{B}}_{1}}\cdot {{\vec{\sigma }}_{1}}+{{\vec{B}}_{2}}(t)\cdot {{\vec{\sigma }}_{2}}+\vec{D}\cdot ({{\vec{\sigma }}_{1}}\times {{\vec{\sigma }}_{2}}),
\end{equation}
in which $J$ denotes the real spin-spin coupling coefficient. This model is called ferromagnetic for $J<0$ and antiferromagnetic for $J>0$. Moreover, $\vec{\sigma}_{j}=(\sigma^{x}_{j},\sigma^{y}_{j},\sigma^{z}_{j}) $  where  $\sigma^{i}_{j} (i=x,y ,z)$ represent the Pauli operaters of  subsystem $j (j=1,2)$ and $ \vec{D}\cdot ({{\vec{\sigma }}_{1}}\times {{\vec{\sigma }}_{2}}) $ is the DM interaction.  Besides, $ \vec{B}_{j} $ $ (j=1,2) $ denotes the magnetic field on site $ j $. We
assume that $\vec{B}_{1}=B_{1}\hat{z}$ and $\vec{B}_{2}=B_{2}\hat{n}(t)$ with the unit vector    $\hat{n}(t)=\left(\sin \theta \cos\varphi(t),\sin \theta \sin\varphi(t),\cos \theta\right)$ where the  azimuthal angle $ \varphi(t) $ changes adiabatically such that condition (\ref{Adiabaticcondition}) is satisfied. Thus, spin 1 is coupled to the static magnetic field while spin 2 is driven  by the adiabatically rotating magnetic field. 
\par  Choosing basis $\{\left|\uparrow\uparrow\right\rangle, \left|\uparrow\downarrow\right\rangle, \left|\downarrow\uparrow\right\rangle, \left|\downarrow\downarrow\right\rangle\}$ and  assuming that the DM coupling is along  $ z $
axis (i.e., $ \vec{D} = D\hat{z} $), we obtain the matrix representation of the Hamiltonian  as follows:
  
\begin{equation}\label{key3}
 H(t)=\left(\begin{array}{cccc}
 B_{1}+B_{2} cos(\theta) & B_{2} sin(\theta)e^{-i\varphi(t)} & 0 & 0 \\
 B_{2} sin(\theta)e^{i\varphi(t)} & B_{1}-B_{2} cos(\theta) &d & 0 \\
 0 & d^{*} &-B_{1}+B_{2} cos(\theta) & B_{2}sin(\theta)e^{-i\varphi(t)} \\
 0 & 0 & B_{2}sin(\theta)e^{i\varphi(t)} & B_{1}-B_{2} cos(\theta) \\
\end{array}\right),
\end{equation}
where $ d\equiv 2(J+iD) $. After lengthy calculation, we find that the eigenvalues and eigenvectors of $ H(t) $, respectively, are given by:
\begin{eqnarray}\label{key6}
\nonumber \xi_{1}&=&	\frac{1}{2}\sqrt{2{{\left|d \right|}^{2}}+4{{B}_{1}}^{2}+4{{B}_{2}}^{2}+2{{\left| \xi  \right|}^{2}}} ,\\
 \xi_{2}&=&-\frac{1}{2}\sqrt{2{{\left|d\right|}^{2}}+4{{B}_{1}}^{2}+4{{B}_{2}}^{2}+2{{\left|\xi  \right|}^{2}}}=-{{\xi}_{1}},\\\nonumber
 \xi_{3}&=&\frac{1}{2}\sqrt{2{{\left|d \right|}^{2}}+4{{B}_{1}}^{2}+4{{B}_{2}}^{2}-2{{\left| \xi  \right|}^{2}}} , \\\nonumber
 \xi_{4}&=&-\frac{1}{2}\sqrt{2{{\left|d \right|}^{2}}+4{{B}_{1}}^{2}+4{{B}_{2}}^{2}-2{{\left| \xi  \right|}^{2}}}=-\xi_{3},
\end{eqnarray}
and

\begin{equation}\label{EigenVector}
\left|\xi_{j}(t)\right\rangle=N_{j}\left(\begin{array}{cccc}
f_{j}e^{-2i\phi(t)} \\
m_{j}e^{-i\phi(t)} \\
g_{j}e^{-i\phi(t)} \\
1 \\
\end{array}\right);
\hspace{1in}
                       \left            (j=1,2,3,4\right),
\end{equation}
where the   time-independent  variables $ f_{j}, m_{j} $, and $ g_{j} $ are presented in the Appendix and  $ N_{j} $ is the normalization coefficient of the jth eigenstate. 
We see that although the eigenvalues are time-independent,
 the eigenstates are clearly time-dependent. 
 
 \section{Adiabatic quantum estimating the direction of the rotating magnetic field \label{QEstimation}}
 In this section, we analyze the bahaviour of the QFI associated with the azimuthal orientation of the rotating magnetic field. Starting from $ \left|\xi_{j}(0)\right\rangle $ and assuming that  the field is rotating adiabatically such that condition (\ref{Adiabaticcondition}) is satisfied, we find that $\left|\psi(t)\right\rangle\cong \left|\xi_{j}(t)\right\rangle  $ up to a calculable phase not affecting the QFI computation. 
 \subsection{Two-qubit probe \label{Two probe}}
 Using both qubits for probing the  azimuthal orientation $ \varphi $ of the rotating magnetic field, we find that the associated QFI corresponding to the jth adiabatic state can be computed using Eq. (\ref{PUREQFI}) as follows
 \begin{equation}
QF{{I}_{j}}=4{{\left| {{N}_{j}} \right|}^{2}}\left( {{\left| {{g}_{j}} \right|}^{2}}+{{\left| {{m}_{j}} \right|}^{2}}+4{{\left| {{f}_{j}} \right|}^{2}}-{{\left| {{N}_{j}} \right|}^{2}}{{\left( {{\left| {{g}_{j}} \right|}^{2}}+{{\left| {{m}_{j}} \right|}^{2}}+2{{\left| {{f}_{j}} \right|}^{2}} \right)}^{2}} \right).
 \end{equation}
 \begin{figure}[ht!]
   \subfigure[]{\includegraphics[width=7cm]{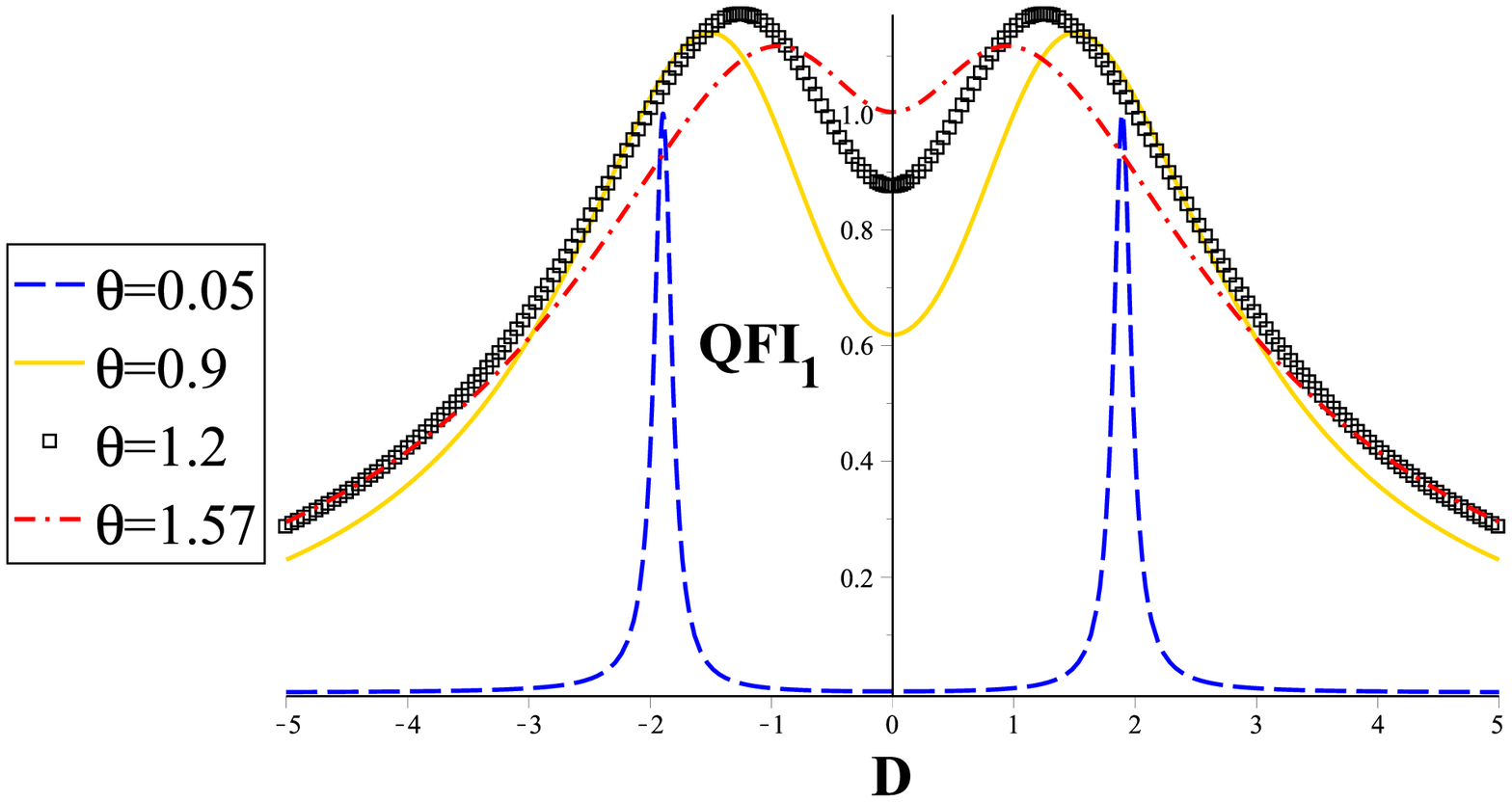}\label{theta1} }
   \subfigure[]{\includegraphics[width=7cm]{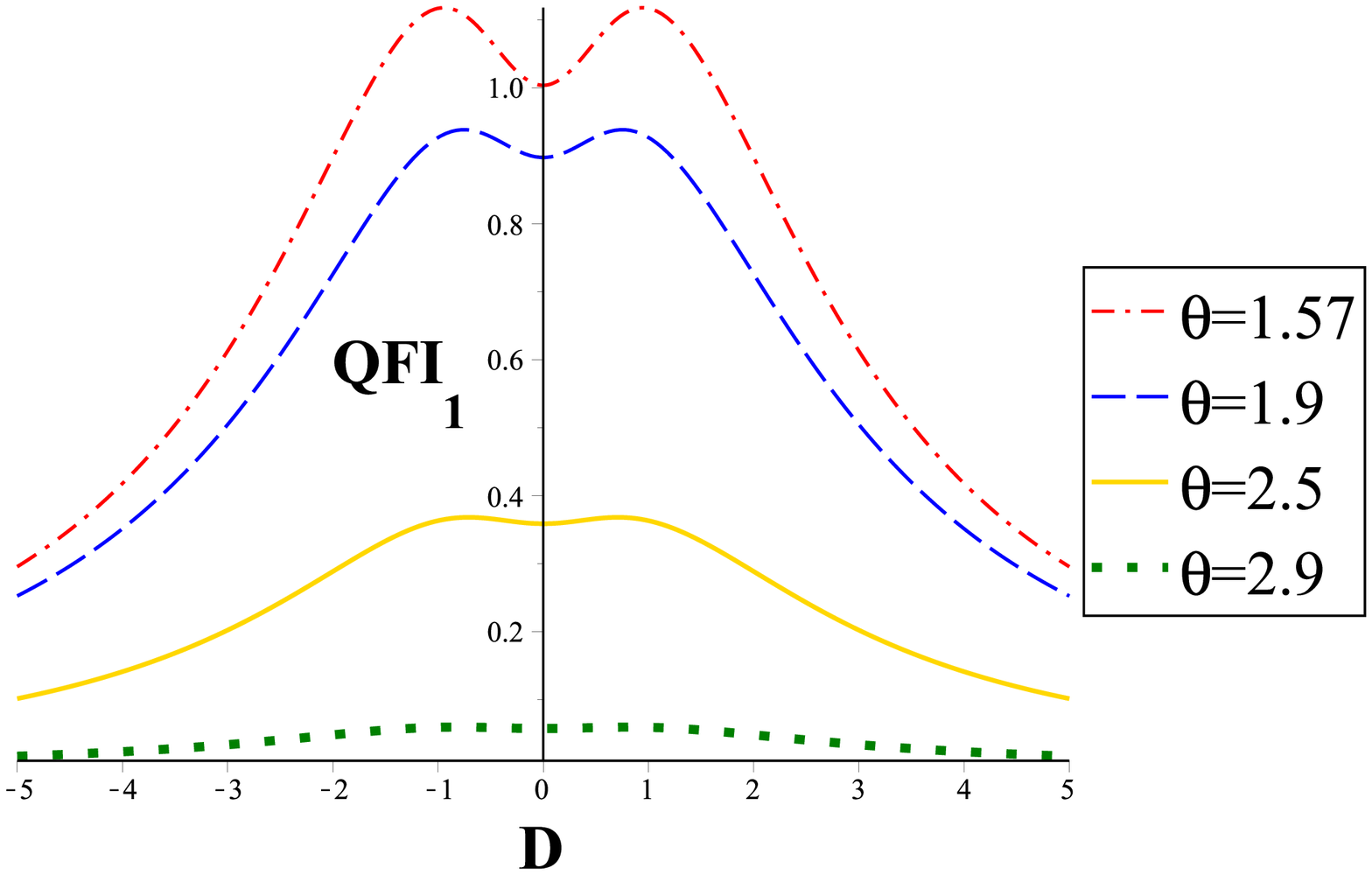}\label{theta2} }
    \caption{The two-qubit QFI, associated with estimating the  azimuthal direction of the rotating magnetic field, versus $ D $  for $ J = 0.1, B_{2} = 3, B_{1} = 1.2 $ and different values of $ \theta $ lying in  range (a)  $[0,\pi/2] $ and (b) $  [\pi/2,\pi]$. } \label{figtheta}
      \end{figure}
      \begin{figure}[ht!]
         \subfigure[]{\includegraphics[width=7cm]{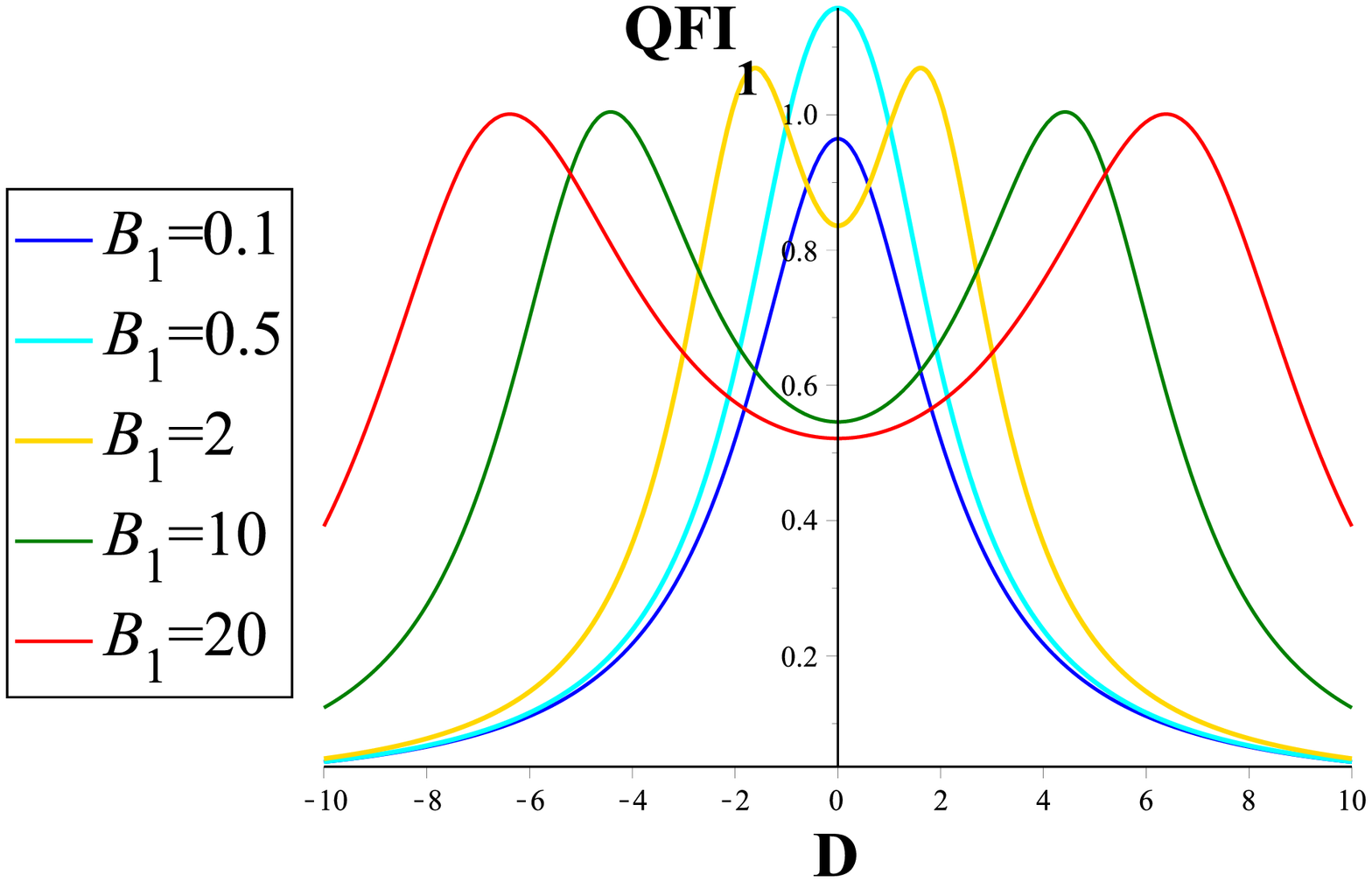}\label{QFIB1} }
         \subfigure[]{\includegraphics[width=7cm]{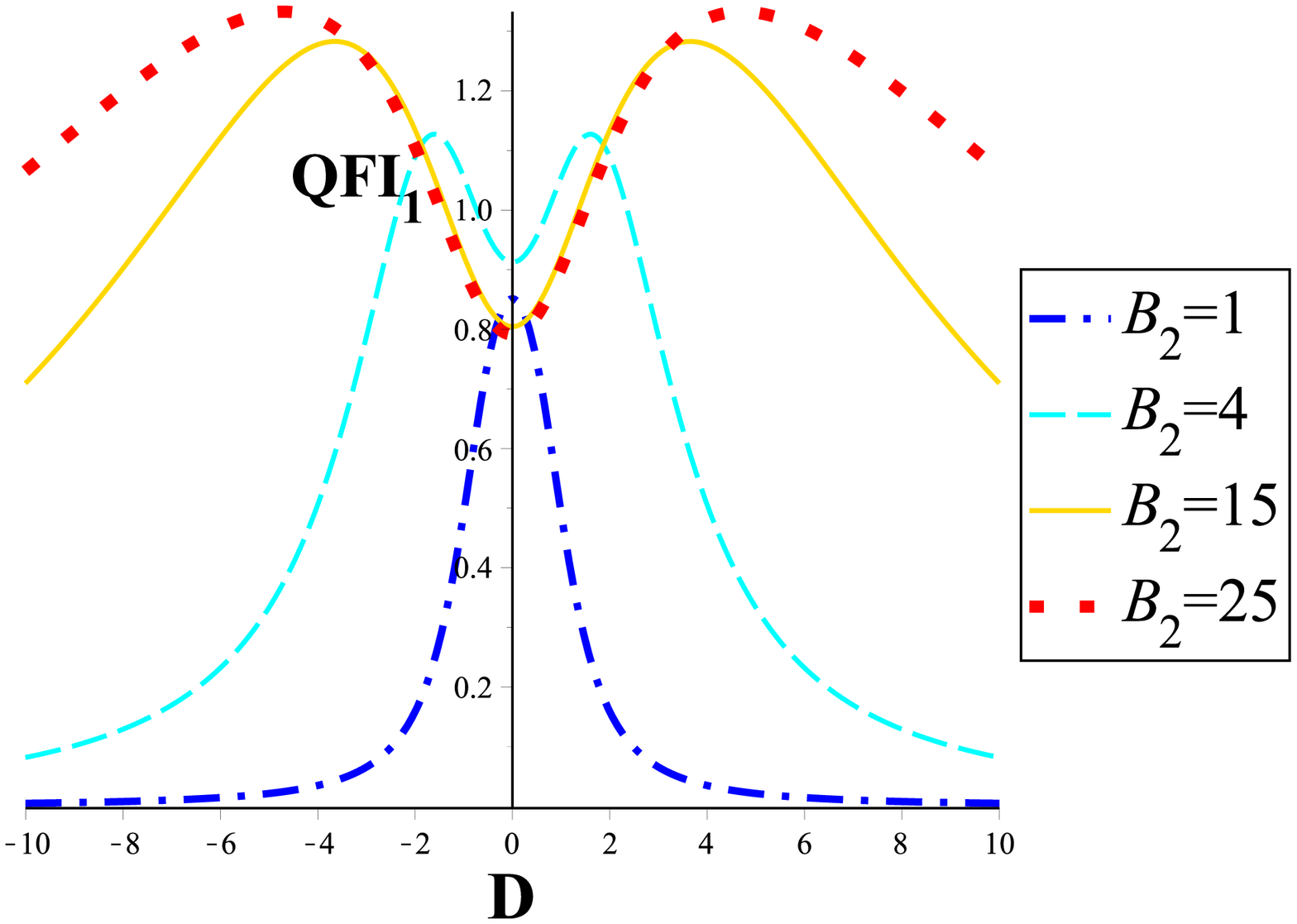}\label{QFIB2} }
          \caption{The two-qubit QFI, associated with estimating the  azimuthal direction of the rotating magnetic field, versus $ D $  for $ J = 1.3, \theta = \pi/4, B_{2} = 3~(B_{1} = 1.5)$   and different values of $ B_{1} (B_{2} ) $  } \label{figB1B2}
            \end{figure}

Our numerical calculation shows that for all values of $J,~D,~B_{1},~ B_{2}  $, and $ \theta $ the QFI of the first (third) adiabatic state is approximately  equal to the QFI of the second (fourth) one,   i.e., $ QF{{I}_{1}}\cong QF{{I}_{2}}$, and $QF{{I}_{3}}\cong QF{{I}_{4}}  $ such that $ | QF{{I}_{1}}- QF{{I}_{2}}|<<1,$  and $|QF{{I}_{3}}- QF{{I}_{4}}|<<1  $. Moreover, in most cases, all of which qualitatively exhibit the same behaviour, and hence we  focus only on the analyses of the behaviour of $ QF{{I}_{1}} $. In addition, it is found that 
\begin{equation}\label{relQFI}
QF{{I}_{j}}(J,D)=QF{{I}_{j}}(D,J);~QF{{I}_{j}}(-J)=QF{{I}_{j}}(J);~QF{{I}_{j}}(-D)=QF{{I}_{j}}(D), 
\end{equation}
denoting that the QFI remains invariant under exchanging the values of $ J $ and $ D $, or reversing their signs.
\par
 Figure \ref{figtheta} illustrates the QFI corresponding to the azimuthal orientation of the rotating field versus the DM coupling for different values of its polar orientation.  We observe that when $ \theta $ lies in range $[0,\pi/2]  $, increasing it causes achievement of the QFI optimum values with weaker DM  coupling. However, if $ \theta $ lies in range $[\pi/2,\pi]  $, increasing it suppresses the QFI and hence the  parameter  estimation  becomes more
 inaccurate.
 
\begin{figure}[ht!]
	\subfigure[]{\includegraphics[width=7cm]{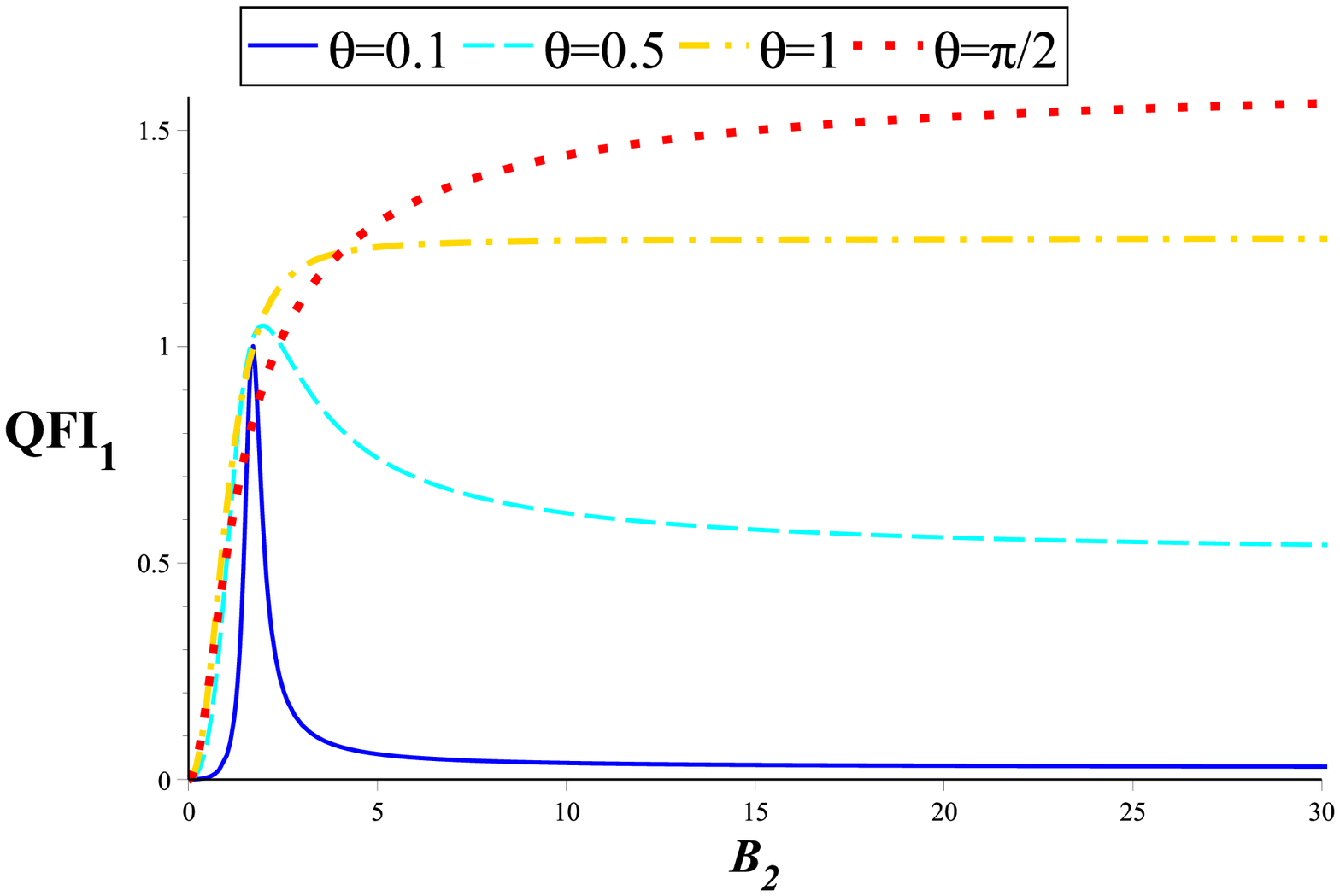}\label{QFI1B2theta} }
	\subfigure[]{\includegraphics[width=7cm]{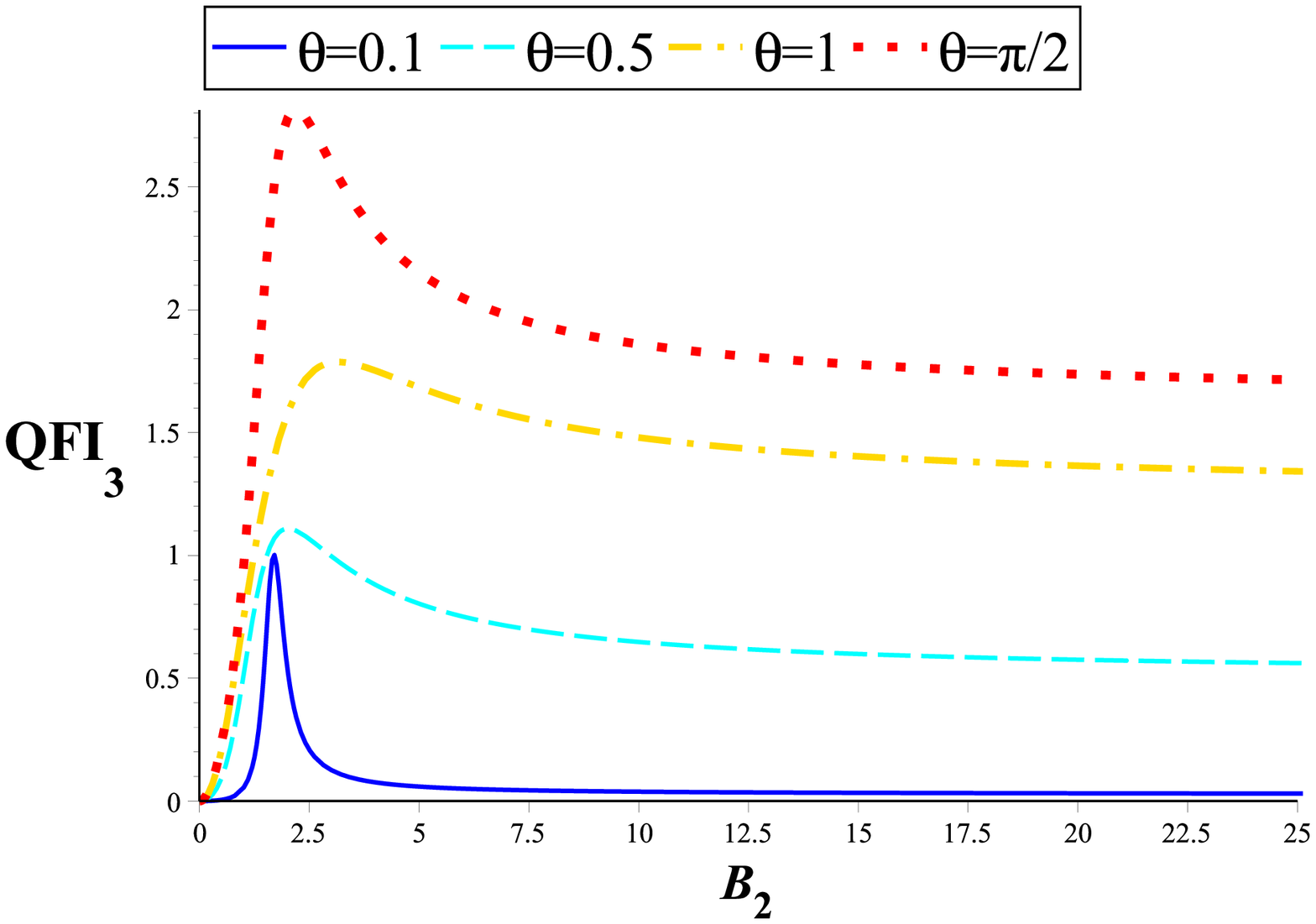}\label{QFI3B2theta} }
	\caption{The two-qubit QFIs corresponding to  adiabatic states (a) 1 and (b) 2, versus $ B_{2} $ for $ J = 1.3,  B_{1} = 1,~ D=0$   and  different values of $ \theta $.  } \label{QFIB2theta}
\end{figure}
The QFI variations versus $ D $ for different intensities of the magnetic fields  are plotted in Fig. \ref{figB1B2}. It is found  that when  $ \theta $ lies in range $[0,\pi/2]~([\pi/2,\pi]) $, weakening the static field, along the positive (negative) direction of z axis, we can achieve the optimal estimation with  weaker DM  coupling (see Fig. \ref{QFIB1}). In particular, applying a weak static magnetic field interacting with the first qubit leads to optimal estimation of the azimuthal angle for $ D \approx 0 $. On the other hand, as seen in Fig. \ref{QFIB2}, intensifying the rotating magnetic field may enhance the estimation. In particular, it leads to  achievement of the optimal values of the QFI for stronger DM coupling. The more exact behavior of the QFI with respect to $ B_{2} $ and the polar orientation of the rotating field for $ D=0 $ can be extracted from Fig. \ref{QFIB2theta}.  We find  that  when $ \theta $ lies in range $ [0,\pi/2] $, $ \text{QFI}_{1,2} $ may be enhanced with an increase in  $ \theta $ (see Fig. \ref{QFI1B2theta}). Moreover, the $ \text{QFI}_{1,2} $ \textit{trapping} occurs  if the qubit is subjected to a strong rotating field. In particular, the $ \text{QFI}_{1,2} $ is  saturated for $\theta=\pi/2  $ and after a certain  value of  $B_2$. In this case, the behavior of $ \text{QFI}_{3,4} $ is slightly different from that of $ \text{QFI}_{1,2} $. In fact, as seen in Fig. \ref{QFI3B2theta}, when $ \theta $ lies in range $ [0,\pi/2] $, $ \text{QFI}_{3,4} $ is enhanced with increasing $ \theta $. Moreover, $ \text{QFI}_{3,4} $ trapping occurs  when the qubit is subjected to a strong rotating magnetic field.  For $D=0$, the $\text{QFI}$ may increase initially and then  decrease as $B_2$ continuously increases. In fact,  intensifying $ \vec{B}_{2} $ can help the two-qubit probe to encode  more information
about the  azimuthal direction of the
rotating magnetic field. However, the rotating field can also lead to more flow of the information from the probes to the environment, because the rotating field simultaneously plays the role of noise in the process of the estimation.
These two mechanism compete with each other and when one of which overcomes the other, the QFI behavior may change. Although at
first the docoherence originated from the noise does not have a dominant effect,  its destructive influence may appear with increasing $ B_{2} $.

 \begin{figure}[ht]
                             \includegraphics[width=12cm]{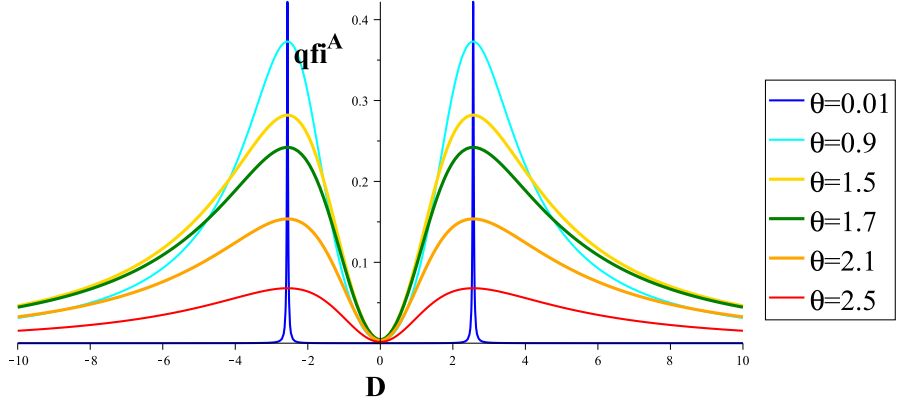}
                             \caption{\small (Color online)   The one-qubit QFI, associated with estimating the  azimuthal angle, versus $ D $  for  $ J = 0.2, B_{1} = 3, B_{2} = 2.2$   and different values of $ \theta $.}
                             \label{qfiAtheta}
                               \end{figure}
                                \begin{figure}[ht]
                                      \includegraphics[width=12cm]{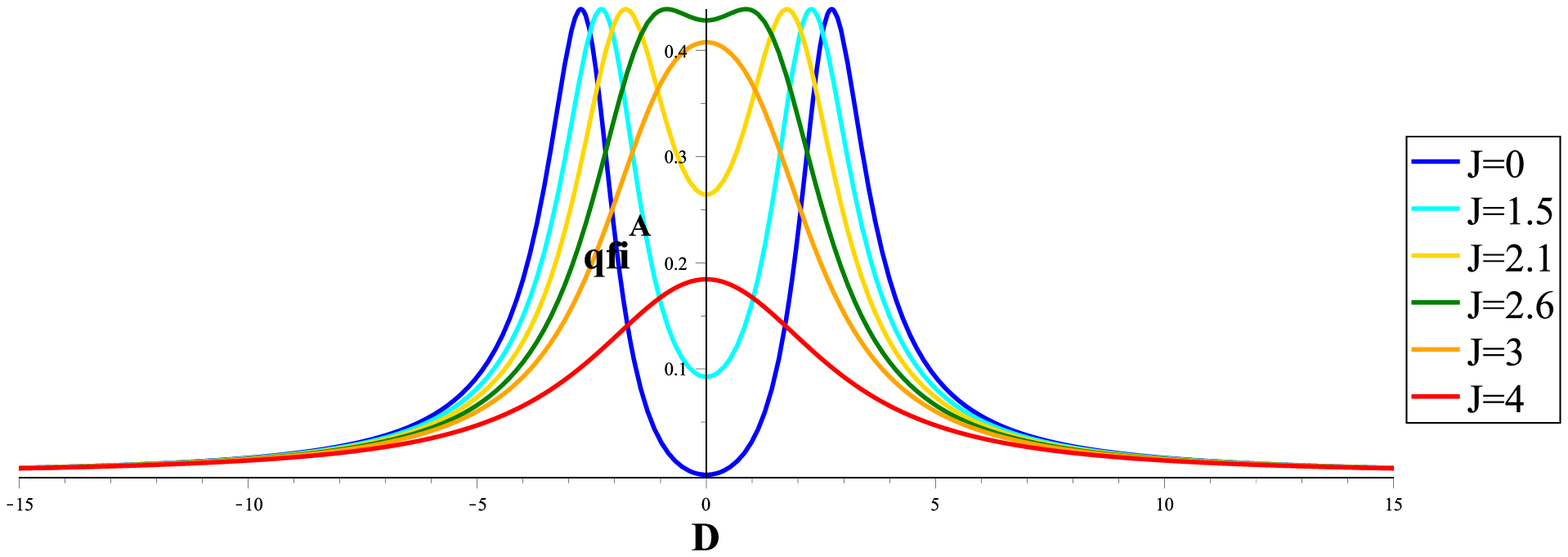}
                                                            \caption{\small (Color online)   The one-qubit QFI, associated with estimating the  azimuthal angle, versus $ D $  for $ \theta = 0.5, B_{1} = 3, B_{2} = 2.5$ and different values of $ J $.}
                                                            \label{qfiAthetaJ}
                                                            \end{figure}

 \subsection{one-qubit probe \label{one probe}}
                       \begin{figure}[ht!]
                                      \subfigure[]{\includegraphics[width=7cm]{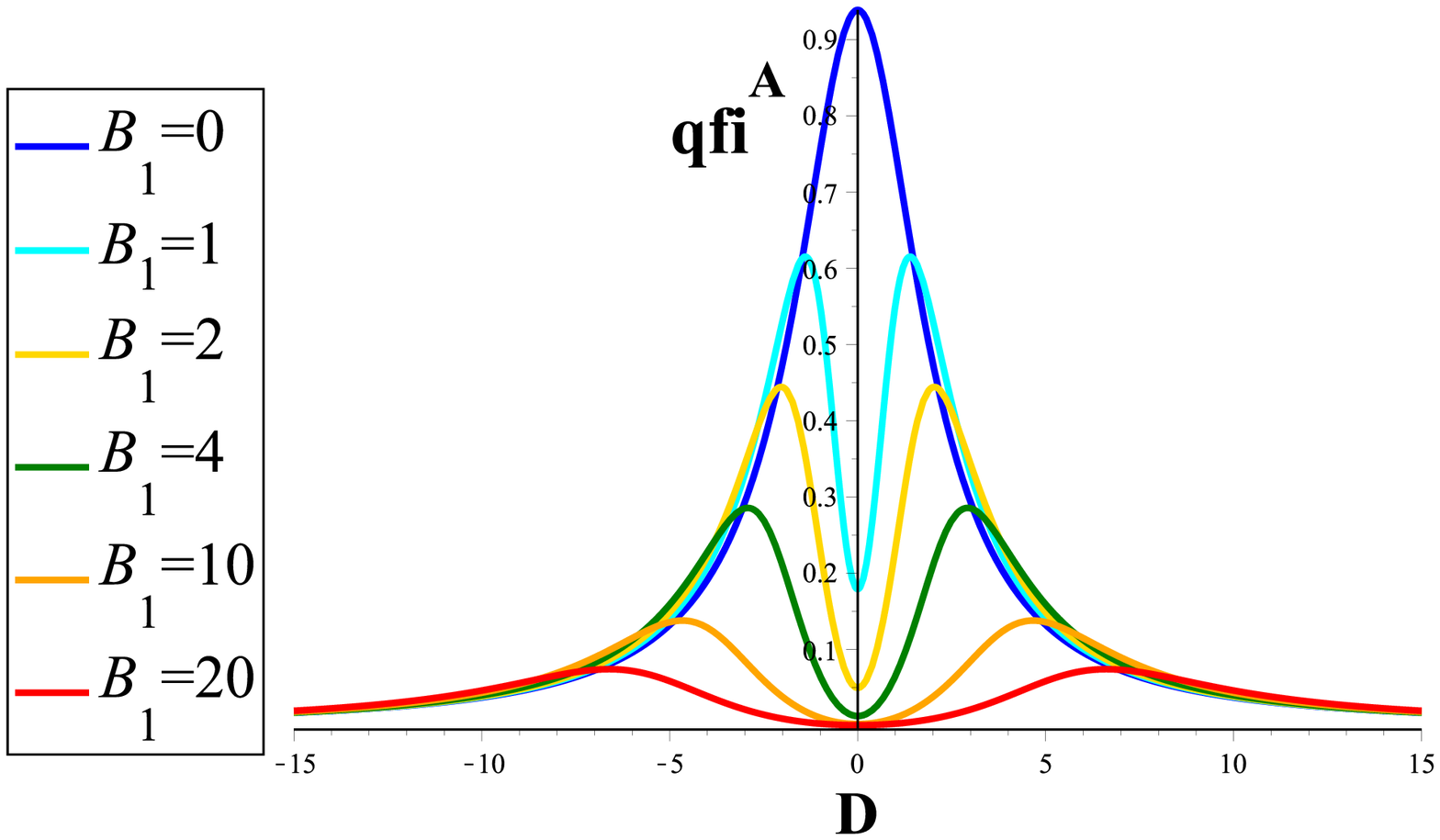}\label{qfiAb1} }
                                       \subfigure[]{\includegraphics[width=7cm]{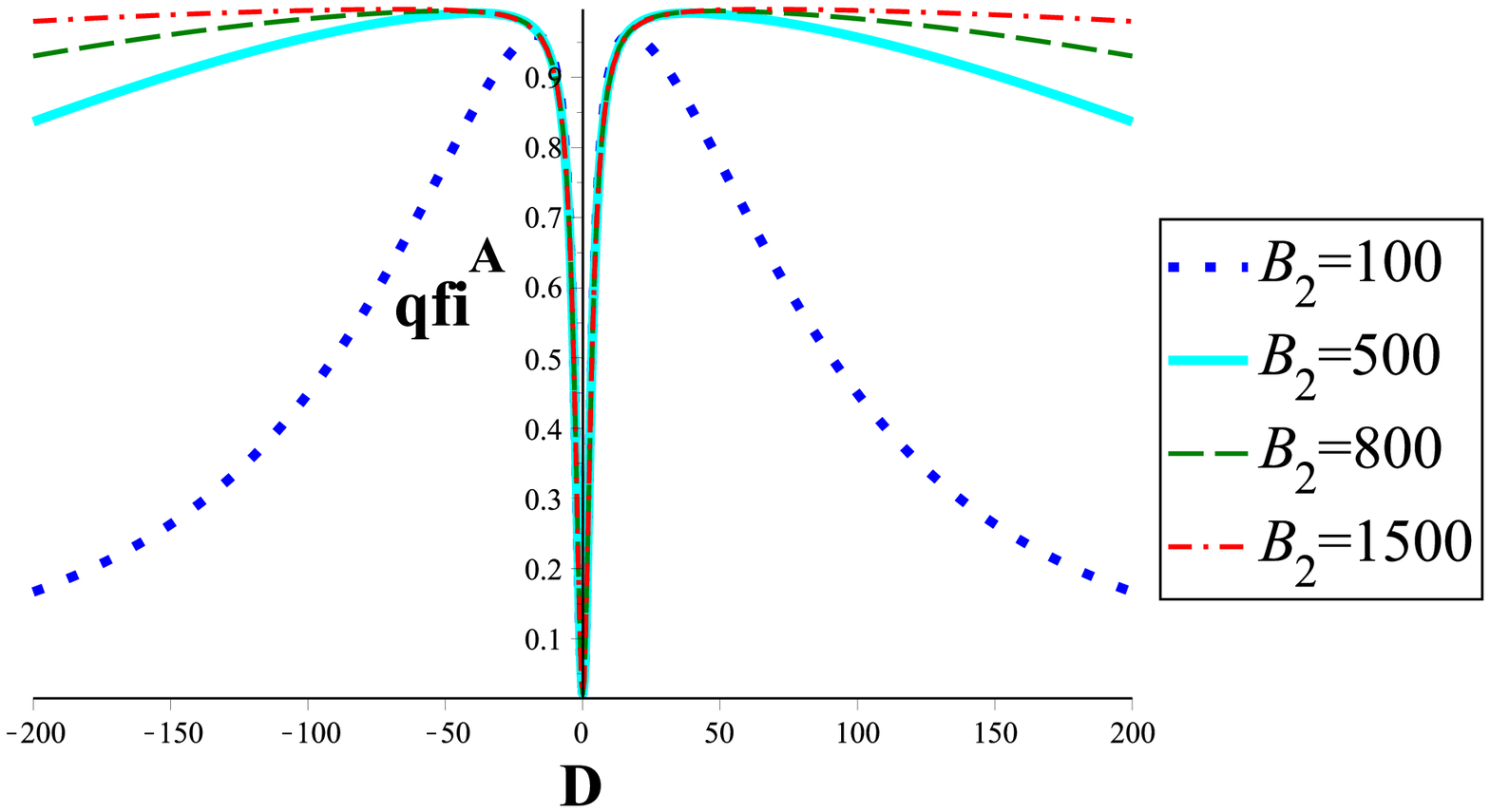}\label{qfiAb2} }
                                     \caption{(Color online) The one-qubit QFI, associated with estimating the  azimuthal direction of the rotating magnetic field, versus $ D $  for $ J = 0.5, \theta = 1.1, B_{2} = 2.2~( B_{1} = 3)$   and different values of $ B_{1} (B_{2} ) $  } \label{qfiAb1b2}
                    \end{figure} 
                
                \begin{figure}[ht]
                	\includegraphics[width=12cm]{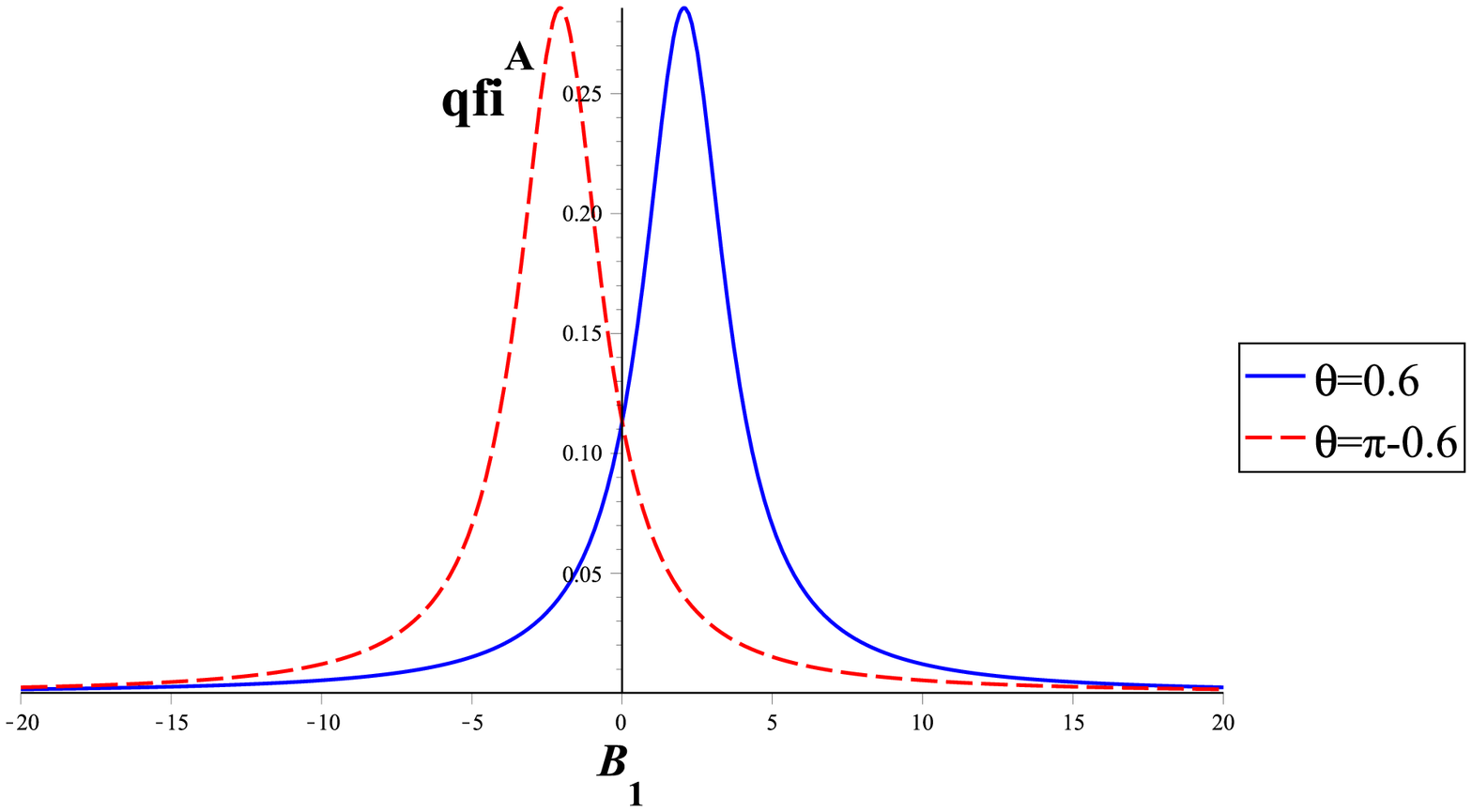}
                	\caption{\small  The one-qubit QFI, associated with estimating the  azimuthal angle, as a function of  $ B_{1} $  for  $ J = 1.3, B_{2} = 1, D = -0.9$   and two values of $ \theta $.}
                	\label{qfiAb1theta}
                \end{figure}

It is  interesting to estimate the azimuthal direction of the rotating magnetic field adiabatically interacting with the second qubit,   using the first spin (qubit A)  as the quantum probe. Applying the adiabatic approximation, we find that the evolution of qubit A is described by the following reduced density matrix obtained by  tracing the adiabatic states on  qubit B:
\begin{equation}\label{rdensity}
{{\rho }^{A}_{j}}(t)=\text{T}{{\text{r}}_{B}}(\left| {{\xi }_{j}} \right\rangle \left\langle  {{\xi }_{j}} \right|)={{\left| {{N}_{j}} \right|}^{2}}\left( \begin{matrix}
   {{\left| {{f}_{j}} \right|}^{2}}+{{\left| {{m}_{j}} \right|}^{2}} & \left( {{f}_{j}}{{g}^{*}_{j}}+{{m}_{j}} \right){{e}^{-i\varphi(t) }}  \\
   \left( {{g}_{j}}{{f}^{*}_{j}}+{{m}^{*}_{j}} \right){{e}^{i\varphi(t) }} & {{\left| {{g}_{j}} \right|}^{2}}+1  \\
\end{matrix} \right).
\end{equation}
Using Eqs. (\ref{02}) and (\ref{rdensity}), one can find  the following analytical expression for the QFI with respect to the azimuthal direction of the rotating field

\begin{equation}\label{rQFI}
\text{qfi}^{A}_{j}=4{{\left| {{N}_{j}} \right|}^{4}}{{\left| {{m}_{j}}+{{f}_{j}}{{g}^{*}_{j}} \right|}^{2}},
\end{equation}
Our numerical computation shows that the QFIs associated with different adiabatic states are approximately equal, i.e.,

\begin{equation}\label{rQFI2}
\text{qfi}^{A}_{1}\simeq \text{qfi}^{A}_{2}\simeq \text{qfi}^{A}_{3}\simeq \text{qfi}^{A}_{4}.
\end{equation}
Moreover, relations similar to (\ref{relQFI}),  extracted for two-qubit QFI, can be presented for one-qubit scenario.

 As plotted in Fig. \ref{qfiAtheta}, it is seen that  a decrease in  $ \theta $ 
does not shift the optimal point of the QFI versus D. Moreover, decreasing  polar angle $ \theta $  raises the optimal value of the
QFI, and hence enhances the optimal precision of estimating the azimuthal direction of the rotating field. 

Figure \ref{qfiAthetaJ} illustrates the effects of the spin-spin coupling on the parameter estimation. We see that although in strong coupling regime an increase in spin-spin coupling may suppress the QFI, in weak DM interaction regime,  strengthening the coupling leads to achievement of the optimal values of the estimation with weaker DM interaction.

In spite of the fact that in two-qubit scenario an increase in the intensity of the static field affecting the first qubit does not considerably decrease the optimal value of the QFI, in one-qubit scenario intensification of $ B_{1} $  leads to the suppression  of the
QFI, thus reducing the optimal
precision of estimation (see Fig. \ref{qfiAb1}). Nevertheless, as demonstrated in Fig. \ref{qfiAb2},  intensifying the rotating field may enhance notably the precision of the estimation. This figure illustrates the asymptotic behavior  of  $ \text{qfi} ^{A}$ with respect to $ B_{2} $, informing the experimentalists how much the rotating magnetic field strength is sufficient to achieve some asymptotically near-optimal QFI values for different values of the DM-interaction. On the other hand, as shown in Fig. \ref{qfiAb1theta}, when the rotating field is weak, the best estimation is achieved by  subjecting the probe to a static magnetic field. However, intensifying $ \vec{B}_{1} $, we see that the parameter estimation  becomes more inaccurate. Moreover, when the rotating magnetic field is strong, applying the static field to the probe   destroys the precision of the parameter estimation and the best estimation is achieved for $ B_{1}= 0 $. In addition, as extracted from Fig. \ref{qfiAb1theta}, the following relations hold:
\begin{equation}\label{onequbitQFIA}
\text{qfi}^{A}_{j}(-B_{1},\theta)=\text{qfi}^{A}_{j}(B_{1},\pi-\theta),~~\text{qfi}^{A}_{j}(B_{1},\theta)=\text{qfi}^{A}_{j}(-B_{1},\pi-\theta).
\end{equation}
Similar result can be obtained for the two-qubit QFI. Therefore, when the rotating field becomes upside down  the QFI can be protected by reversing the direction of the static magnetic field applied at the location of the probe.
\par
Now an important question arises: under what conditions is use of the qubit not affected by the rotating field more efficient for probing the azimuthal orientation of $ \vec{B}_{2} $ than use of the qubit driven by that field? First, it should be noted that the two-qubit scenario for the estimation is always more efficient than the one-qubit one, i.e., $ \text{QFI}\geq \text{qfi} $. On the other hand, the following results have been obtained using numerical calculation;\\
a) when  $ B_{1}=0 $ or $ J\gg 1 $ or $ D \gg 1 $, we find that the information extracted from the first qubit can lead to better estimation of the azimuthal angle, than  the information  achieved from the second one. In fact, under those conditions, we obtain $\text{qfi}^{A}_{j}\geq \text{qfi}^{B}_{j}$; Moreover, if $ J\gg 1 $, the optimal estimation occurs for $ D=0 $.\\
b) If $ J $ or $ D $ equals zero and $ B_{1}=B_{2} $, we find that $\text{qfi}^{B}_{j}\geq \text{qfi}^{A}_{j}$ for $ j=1,2 $ while 
$\text{qfi}^{A}_{j}\geq \text{qfi}^{B}_{j}$ for $ j=3,4 $.

\section{Summary and conclusions \label{conclusion}}
\par 
In this paper, we have investigated the adiabatic estimation of the direction of a  magnetic field from the perspective of the QFI. In fact, we applied two qubits as potential probes, described by the Heisenberg XX model, such that one of
which was driven by the rotating magnetic field, as
adiabatic condition was satisfied, and the other one experienced
a static magnetic field. Adiabatic evolution guarantees continuous evolution of the instantaneous eigenstates of the Hamiltonian at one time  to the
corresponding eigenstates at later times. After analytical computation of the two-qubit QFI associated with the azimuthal orientation of the rotating field, we exactly analyzed the effects of intensities of the magnetic fields, the DM interaction,
and the coupling coefficient $ J  $ on the optimal estimation. In particular, we found that the polar orientation of the rotating magnetic field plays a key role in the process of estimating its azimuthal direction such that   when $ \theta $ lies in range $[0,\pi/2]  $, increasing the polar angle causes achievement of the QFI optimum values with weaker DM  coupling. However, if $ \theta $ lies in range $[\pi/2,\pi]  $, increasing it suppresses the QFI and hence the  parameter  estimation  becomes more
 inaccurate.
\par
We also discussed how  the azimuthal direction of the rotating field can be estimated  using the qubit not affected by that field. The one-qubit QFI 
was computed for this purpose and its behaviour was investigated in detail.
In particular, we showed that
when the rotating field is weak, the optimal estimation is
achieved by subjecting the probe to a static magnetic field.
  Moreover, we investigated under what conditions the use of the qubit not affected by the rotating field is more efficient for the estimation than the use of the qubit driven by that field.

\par

\noindent\textbf{Data accessibility.} This paper does not have any experimental data.

\noindent\textbf{Competing interests.} We have no competing interests.

\noindent\textbf{Authors' contributions.}
All the authors conceived the work and agreed on the approach to pursue. H.R.  planned and supervised the project, carried out  the theoretical calculations and wrote the manuscript.  L.F.-Sh., H.R.,  and M.G. extracted  and    discussed the results. All authors gave final approval for publication.

\noindent\textbf{Funding statement. }
 H.R. acknowledges funding by the grant no. 3539HRJ of Jahrom University.

\appendix
\section{}
The time-independent coefficients appeared in (\ref{EigenVector}) are given by:
\begin{eqnarray}\label{coefficients}
\nonumber f_{j}&=&	\dfrac{-2\,B^{2}_{{2}}d \sin^{2} \theta }{P_{j}},\\\nonumber
 m_{j}&=&\dfrac{2\,B_{{2}}\sin \theta~ d \left( B_{{2}}\cos \theta~ +B_{{1}}-\xi_{{j}} \right)
  }{P_{j}},\\
g_{1,2}&=&-\dfrac{B_{{2}}\sin \theta \left( 4B^{2}_{1}-4B_{1}\xi_{1,2}+|d|^{2}+|\xi|^{2} \right)
  }{P_{1,2}},\\\nonumber
   g_{3,4}&=&-\dfrac{B_{{2}}\sin \theta \left( 4B^{2}_{1}-4B_{1}\xi_{3,4}+|d|^{2}-|\xi|^{2} \right)
      }{P_{3,4}},\nonumber
\end{eqnarray}

 where
\begin{multline}
 P_{1,2}=4\,B_{{2}} \left( \dfrac{  \left| \xi \right|   ^{2}}{4} -\dfrac{  \left| d \right|   ^{2}}{4}+B_{{1}} \left( B_{{1}}-\xi_{{1,2
}} \right)  \right) \cos \theta + ( B_{{1}}-\xi_{{
1,2}} )  \big(  | \xi|   ^{2}- | d |   ^{2}\big)+4\,B_{{1}}{
B^{2}_{{2}}},
\end{multline}
\begin{multline}
 P_{3,4}=4\,B_{{2}} \left( \dfrac{ - \left| \xi \right|   ^{2}}{4} -\dfrac{  \left| d \right|   ^{2}}{4}+B_{{1}} \left( B_{{1}}-\xi_{{3,4
}} \right)  \right) \cos \theta+ (- B_{{1}}+\xi_{{
3,4}} )  \big(  | \xi|   ^{2}+ | d |   ^{2}\big)+4\,B_{{1}}{
B^{2}_{{2}}},
\end{multline}
and where
 \begin{equation}
 |\xi|^{2}=\sqrt {-4\,  \cos^{2}  \theta    
 \left| d \right|  ^{2}{B_{{2}}}^{2}-8\,\cos \theta
 ~  \left| d \right|   ^{2}B_{{1}}B_{{2}}+
  \left| d \right|   ^{4}+4\,   \left| d \right| 
  ^{2}{B^{2}_{{2}}}+16\,{B^{2}_{{1}}}{B^{2}_{{2}}}}.
 \end{equation}

\end{document}